\documentclass[prl,aps,twocolumn,preprintnumbers,showpacs]{revtex4}

\usepackage{amsmath,amssymb,bm}
\usepackage{graphicx}
\usepackage{color}

\begin{document}
\title{
Quantum criticality in a Mott pn-junction in an armchair carbon nanotube}
\author{Wei Chen$^1$, A. V. Andreev$^1$, L. I. Glazman$^2$}
\affiliation{1.Department of Physics, University of Washington, Seattle, WA
98195-1560, USA}

\affiliation{2.Department of Physics, Yale University, New Haven, CT 06520, USA}

\begin{abstract}
In an armchair carbon nanotube pn junction the p- and n- regions are separated by a region of a Mott insulator, which can backscatter electrons only in pairs. We predict a quantum-critical behavior in such a pn junction. Depending on the junction's built-in electric field $E$, its
conductance $G$ scales either to zero or to the ideal value $G=4e^2/h$ as the temperature $T$ is lowered. The two types of the $G(T)$ dependence indicate the existence, at some special value of $E$, of an intermediate quantum critical point with a finite conductance $G<4e^2/h$. This makes the pn junction drastically different from  a simple barrier in a Luttinger liquid.
\end{abstract}
\maketitle

Transport measurements in carbon nanotube (CNT) devices reveal strongly correlated behavior of conduction electrons. Experiments on tunneling into single-wall CNT~\cite{McEuen1,Dekker} or across a barrier interrupting a CNT~\cite{Dekker} demonstrated a power law bias- and temperature- dependence of the current, consistent with a gapless excitation spectrum and Luttinger correlations of the electron liquid. Recent experiments~\cite{Bockrathgap} showed that armchair CNTs develop a gap in the spectrum of charge excitations at zero doping.
The gap formation could be attributed to the electron Umklapp
processes~\cite{Odintsov1999,Nersesyan2003}, which back-scatter pairs of electrons. These processes drive armchair CNTs into the Mott-insulating state.  In contrast, undoped semiconducting CNTs are simple band insulators.  Although  Mott and  band insulators are qualitatively different, no clear experimental signature of that dichotomy has been observed  in CNTs.

In this Letter we predict that the difference should manifest itself in the conduction of a pn junction formed by bipolar doping of a CNT. In contrast to a simple ``band'' pn junction, in which the barrier between the p- and n- regions is formed by a band insulator, a ``Mott'' pn junction formed in a pristine armchair CNT does not back-scatter single electrons. However, the Umklapp backscattering of pairs of electrons remains effective near its center. In short junctions these processes are irrelevant and do not alter the perfect zero-temperature conductance ($G=4e^2/h$). In longer junctions, Umklapp processes bring about the Mott-insulator state and drive the zero-temperature conductance to zero. The bipolar doping of a CNT can be achieved by gating the nanotube~\cite{McEuen_pn}, which enables one to change the junction length by controlling the built-in electric field $E$. By varying $E$ at a fixed low temperature $T$, one can tune the conductance $G$ of the junction between almost perfect and zero values. Lowering the temperature makes this crossover sharper and culminates in a zero-temperature quantum phase transition occurring at some critical field value $E=E^*$. The described dependence on $E$ and $T$
the Mott pn junction conductance, $G(T,E)$, is drastically different from that of a band pn junction, where $G$ vanishes at any $E$ in the low temperature limit, $T \to 0$. We make detailed predictions for the $G(T,E)$ dependence in the vicinity of the quantum phase transition.

\begin{figure}[ptb]
\includegraphics[width=8.0cm]{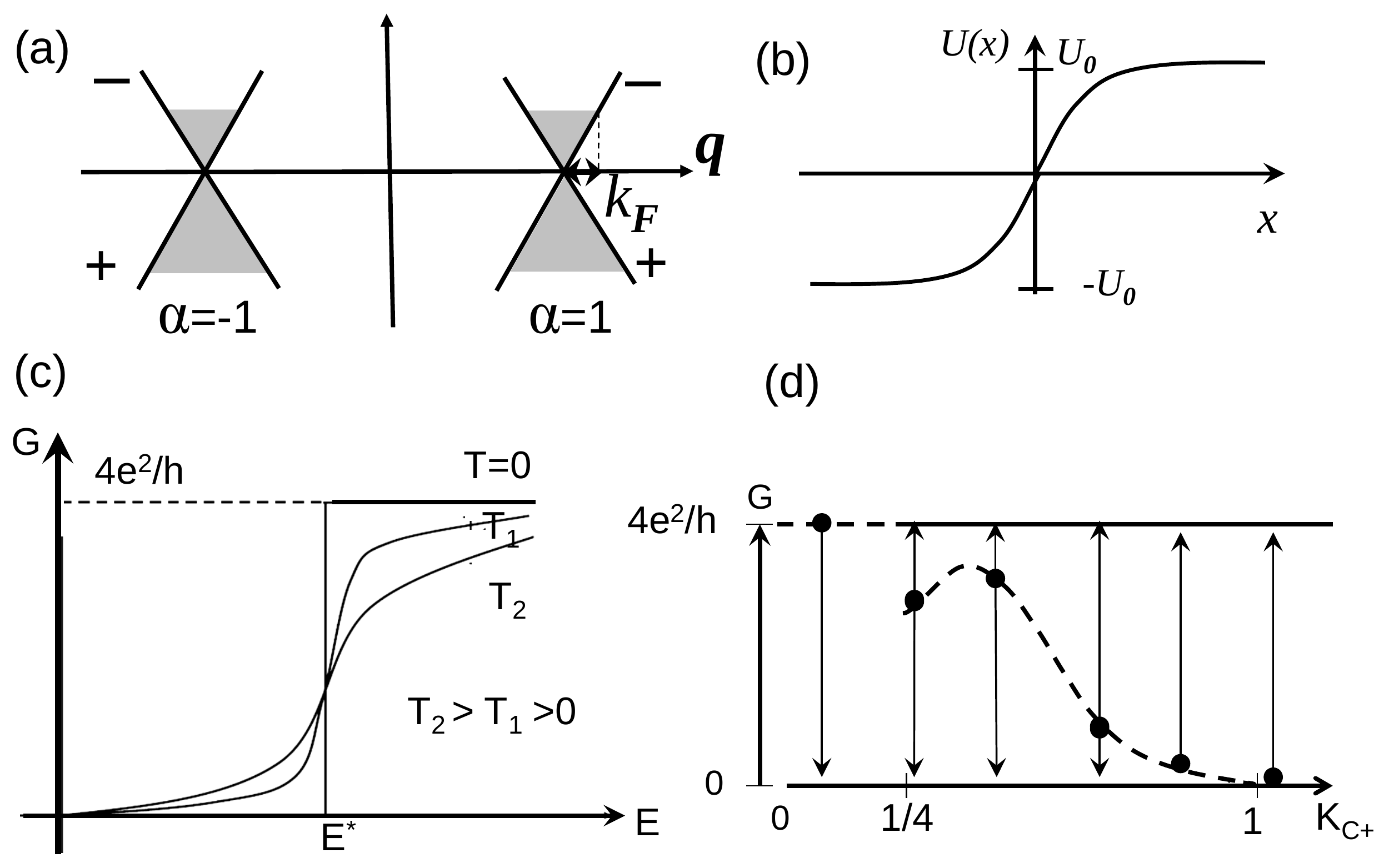}
\caption{(a) Electron spectrum near the Dirac points $\alpha=\pm 1$ ($+$ or $-$ indicates parity). (b) Near $x=0$ the gate potential is $U(x)\approx -e E x$ and saturates to $\pm U_0$ in the p- and n- regions. (c) Schematic picture of the conductance dependence on $E$ at different temperatures. The $G(E)$ curves at different temperatures intersect at the fixed point and increase monotonically with $E$. The step in $G(E)$ at $T=0$, indicates a quantum phase transition. (d) RG flow for the conductance. } \label{fig:pnjunction}
\end{figure}

We consider a single pn-junction that is formed by a static potential $U(x)$ imposed by two gates, as in Fig.~\ref{fig:pnjunction}(b). In the p- and n- regions $U(x)$ saturates to $\pm U_0$, while near the center of the pn-junction $U(x)\approx eE x$, where $-e$ is the electron charge and $E$ is  the built-in electric field. If the gate potential is not too strong,  $|U_0| \ll v_F/R$, where  $R$ is the CNT radius (we set $\hbar=1$ in the intermediate formulas), the pn-junction may be modeled using the low energy CNT band structure shown in Fig.~\ref{fig:pnjunction} (a). Two spin degenerate bands intersect at two Dirac points (valleys, $\alpha=\pm 1$). The electron wave functions in the intersecting bands have opposite parity $P=\pm 1$. They are either symmetric  ($P=1$) or antisymmetric ($P=-1$) in the AB sublattices~\cite{Jishi1993}.  We assume that $U(x)$ is smooth on the interatomic scale and does not break the symmetry between the A and B sublattices. Then both inter- and intra-valley backscattering is absent, and we may write the noninteracting part of the electron Hamiltonian as
\begin{equation}\label{eq:H_0}
H_{0}=\sum_{\alpha r \sigma}\int dx\, \psi_{\alpha r
\sigma}^{+}(x) \left[-i\, r\,v_F\partial_{x}+U(x)\right]\psi_{\alpha r \sigma}(x). \nonumber
\end{equation}
Here $v_F\approx 8 * 10^5$ m$/$s is the Fermi velocity, $\alpha=\pm$ is the valley index, $r=\pm 1$ labels left and right movers, and $\sigma$  the electron spin.

We assume that the number of unit cells $N$ around the circumference of an $(N,N)$ armchair CNT is large. To zeroth order in $1/N$ the Coulomb interaction of electron-electron (e-e) backscattering is absent~\cite{Balents1997}. The forward scattering part of the e-e interactions is,
\begin{equation}\label{eq:V_forward}
\!V_f=\frac{V(0)}{2}\int dx n^2(x); \,\,\, n(x)= \sum_{\alpha r \sigma} \psi_{\alpha r
\sigma}^{+}(x)\psi_{\alpha r
\sigma}(x). \nonumber
\end{equation}
Here $V(0)\approx 2 e^2 \ln ( d/R)$ is the forward scattering matrix element (we assume that the Coulomb interaction is screened by the gate at a distance $d$ from the CNT).

The gate potential $U(x)$ results in a position-dependent doping density characterized by the Fermi wave vector $k_F(x)\sim - U(x)/v_F$. We can bosonize the electron operators by the standard procedure~\cite{Giamarchi}
\begin{equation}\label{eq:bosonization_of fermions}
    \psi_{\alpha r \sigma}(x)=\frac{F_{\alpha \sigma}}{\sqrt{2\pi \xi}}\,
e^{ir\,[\int^x k_F(x') dx' - \, \Phi_{\alpha
\sigma}(x)]+i\Theta_{\alpha \sigma}(x)}.
\end{equation}
Here  $\xi\sim R$ is the short distance cutoff,  $F_{\alpha \sigma}$ are Klein factors. The bosonic fields $\Phi$ and
$\Theta$  obey the standard commutation relations $[\Phi_{\alpha \sigma}(x),\Theta_{\alpha^{\prime}
\sigma^{\prime}}(x^{\prime})]=-i \pi \delta_{\alpha \alpha^{\prime}}
\delta_{\sigma \sigma^{\prime}} \theta(x-x^{\prime})$.

In the bosonic representation the electron density is $n(x)=4 k_F(x)/\pi-\sum_{\alpha \sigma} \partial_x \Phi_{\alpha \sigma} (x)/\pi$ and the forward scattering part of the Hamiltonian $H_f\equiv H_0+V_f$ reduces to the same form as in a uniform CNTs~\cite{CNT_Luttinger}
\begin{equation}\label{eq:H_forward}
H_f=\int \frac{d x}{2\pi} \sum_{j}u_j
\left[ K_{j}(\partial_x \Theta_{j})^2 + (\partial_x \Phi_{j})^2/K_{j}\right].
\end{equation}
Here $j=c\pm, s\pm$ labels charge ($c$) and spin ($s$) modes that are symmetric ($+$) or antisymmetric ($-$) in the valley index $\alpha$. They are related to the
fields $\Phi_{\alpha \sigma}$ by $\Phi_{\alpha \sigma} = \left[ \Phi_{c +} + \alpha \Phi_{c -} + \sigma
\Phi_{s +} +\alpha \sigma \Phi_{s -} \right]/2$.
Only the $c+$ mode carries charge and the other three are neutral.
The mode velocities are  $u_j=v_F/K_j$, where the Luttinger parameters are $K_{c+}=1/\sqrt{1+4V(0)/\pi v_F}\ll1$ and $K_j = 1$ for the neutral modes.

Backscattering interactions are small in $1/N$~\cite{Balents1997}, and may be treated as perturbations to the Hamiltonian (\ref{eq:H_forward}). Some of them are relevant and qualitatively change the low energy physics. The most relevant backscattering interaction corresponds to the so-called Umklapp processes, which scatter two right-movers into left-moving states or vice-versa~\cite{CNT_Luttinger, Odintsov1999}. In the presence of doping the Umklapp Hamiltonian can be written as~\cite{Nersesyan2003}
\begin{eqnarray}\label{eq:H_Umklapp}
 H_{U}&=&- \int \frac{ dx}{2(\pi \xi)^2} \cos{\left(2\Phi_{c+}(x)-4\int^x_0 k_F(y)dy \right)}\times \nonumber \\
               &&\left\{ g_3 \cos[2\Theta_{s-}(x)] + (g_3-g_1)\cos[2\Phi_{s+}(x)] \right.\nonumber \\
                 && \left.+ \, g_1(\cos[2\Phi_{c-}(x)]-
                \cos[2\Phi_{s-}(x)])\right\}.
\end{eqnarray}
Here  the coupling constants $g_1$ and $g_3$ are of order $e^2/N$ and $g_3>g_1$.
The low energy electron Hamiltonian of the pn-junction is given by the sum of Eqs.~(\ref{eq:H_forward}) and (\ref{eq:H_Umklapp}),
\begin{equation}\label{eq:H_full}
    H=H_f+H_U.
\end{equation}
The position-dependent Fermi vector $k_F(x)$ in Eq.~(\ref{eq:H_Umklapp}) saturates to constant values $\pm k_0$ in the n- and p- regions whereas near $x=0$ it has a linear dependence of $x$,
\begin{equation}\label{eq:k_F_L_E}
    k_F(x)=-x/L_E^2.
\end{equation}
Here the length scale $L_E$ is defined by the built-in electric field $E$ controlled by the gate voltages, $L_E\sim \sqrt{v_F/eE}$.

Relating the current operator to charge field as $j= - 2 e \partial_t \Phi_{c+}(x=0, t)/\pi$ and using the Kubo formula we can express the device conductance as
\begin{equation}\label{eq:conductance_Green_function}
    G=i \frac{8e^2}{\pi h}\lim_{\omega \to 0}  \omega \, \mathcal{G}_\omega(0,0).
\end{equation}
Here $\mathcal{G}_\omega(x,x')$ is the retarded Green's function of the charge field,
$\mathcal{G}_\omega(x,x') =-i \int^\infty_0 dt e^{i\omega t} \left\langle [\Phi_{c+}(x, t), \Phi_{c+}(x', 0)]\right\rangle$.

In the absence of Umklapp interactions evaluation of the Green's function is straightforward and gives
\begin{equation}\label{eq:Green_function}
    \mathcal{G}^0_\omega(x-x')=-i \pi (K_{c+}/2 \omega_+)\,  e^{i \omega_+ |x-x'|/u_{c+}},
\end{equation}
where $\omega_+=\omega + i \eta$. When substituting this expression into Eq.~(\ref{eq:conductance_Green_function})
it should be born in mind that the dc conductance is controlled by the leads, where LL effects are absent. Thus the Luttinger parameter  should be set to unity, $K_{c+}\to 1$~\cite{maslov}, yielding $G_0=4 e^2/h$.

The Umklapp processes degrade electric current by backscattering pairs of electrons and thus decrease the device conductance. They are most effective near the zero doping point $x=0$ and strongly suppressed deep in the p- and n- regions. At low temperatures, $T\ll v_F k_0$, the Umklapp backscattering in the p- and n- regions result in exponentially small $\sim \exp(-2v_F k_0/T)$ resistivity. We assume that the length of the p- and n- regions is not sufficient to compensate for this exponential smallness and neglect this contribution. In this approximation backscattering arises from the spatial region $|x| \lesssim T/eE$, where $|k_F(x)| \lesssim T/v_F$. At $T\ll U_0$ the spatial dependence of $k_F(x)$ in this region is linear. Therefore the pn-junction may be modeled by the Hamiltonian (\ref{eq:H_full}) with  $k_F(x)$  given by Eq.~(\ref{eq:k_F_L_E}) in the entire space.

Spatial localization of backscattering enables us to express Green's function of the charge mode as
\begin{equation}\label{eq:Green_function_T_matrix}
    \mathcal{G}_\omega(0,0)=\mathcal{G}^0_\omega(0)+\mathcal{G}^0_\omega(0) \int_{-\infty}^\infty dx dx'T_\omega (x,x') \mathcal{G}^0_\omega(0).
\end{equation}
Here $T_\omega (x,x')$ is the part of the T-matrix that corresponds to scattering of  a plasmon into a single plasmon. Using  Wick's theorem (c.f. Appendix A of Ref.~\cite{Chen2010}) one can show that it can be expressed as
\begin{eqnarray}\label{eq:T_expression}
    T_\omega(x,x')&=&- i \int^\infty_0 dt \, e^{i\omega t} \left\langle [\partial_{\Phi}\mathcal{H}_U(x,t), \partial_{\Phi} \mathcal{H}_U(x',0)]\right\rangle \nonumber \\
    && + \,\delta(x-x') \langle \partial^2_{\Phi}\mathcal{H}_U(x)\rangle,
\end{eqnarray}
where $\mathcal{H}_U(x,t)$ is the Umklapp Hamiltonian density and  $\partial_\Phi$ denotes a partial derivative with respect to $\Phi_{c+}$.

In the low frequency limit needed for evaluating the dc conductance the unperturbed Green's functions $\mathcal{G}^0_\omega$ in Eq.~(\ref{eq:Green_function_T_matrix}) are given by Eq.~(\ref{eq:Green_function}) with $K_{c+}=1$. Then the deviation $\delta G$ of the device conductance from the ideal value, $4e^2/h$, may be expressed as
\begin{equation}\label{eq:delta_G}
    \delta G=- i \frac{e^2}{h} \lim_{\omega \to 0} \frac{2\pi}{\omega} \int_{-\infty}^\infty dx \, dx' T_\omega (x,x').
\end{equation}
The  T-matrix properties are dominated by the fluctuations of bosonic modes with frequencies on the order of the temperature $T$ and characteristic spatial scale of $L_T\sim v_F/T$. Provided the device length is longer than $L_T$ the T-matrix needs to be evaluated using the Hamiltonian (\ref{eq:H_full}) with the values of the Luttinger parameters corresponding to the device interior.

The energy gap $\Delta\sim ( v_F/\xi)  (g_3/v_F)^{1-K_{c+}}$~\cite{Odintsov1999} induced by the Umklapp interaction in a uniform CNT at zero doping defines an additional characteristic length scale, $\zeta = v_F/\Delta$. Backscattering at the pn-junction is weak for $L_E\ll \zeta$ and strong for $L_E\gg \zeta$.

For $L_E\ll \zeta$ the correction to the ideal conductance may be expanded in perturbation series in $H_U$, Eq.~(\ref{eq:H_Umklapp}). Using Eqs.~(\ref{eq:delta_G}), (\ref{eq:T_expression}) we get to second order in $H_U$,
$\delta G^{(2)}\!=- c \left( 2\pi T \xi/u_{c+}\right)^{2K_{c+}} \!(L_E/\xi)^2 (3 g_1^2\!-2 g_1 g_3 + 2g_3^2)/v_F^2$,
where $c=(e^2/h)B(K_{c+}+1, K_{c+}+1)/2\pi$, with  $B(x, y)$ being the Euler Beta function.
This result may be rewritten as $\delta G^{(2)}\sim (e^2/h) (L_E/\zeta)^2 (T/\Delta)^{2K_{c+}}$.
It vanishes at $T\to 0$, which corresponds to irrelevance of point-like Umklapp scattering in the renormalization group (RG) sense.

Although higher order terms in perturbation theory are smaller in powers of $L_E/\zeta$ they have different temperature dependence. In the fourth order correction to the ideal conductance the most relevant term is $\delta G ^{(4)}\sim - (e^2/h)(L_E/\zeta)^4 (T/\Delta)^{8K_{c+}-2}$.

This result can also be obtained using the following RG considerations. Because of localization of backscattering to the  the region of size $|x|\leq L_E$  at low temperatures, $T\ll v_F/L_E$, the pn-junction acts as a point scatterer. In contrast to a \emph{potential} point scatterer~\cite{Kane1992} the pn-junction can only scatter pairs of electrons. The Umklapp scattering may be described by an effective ``impurity'' Hamiltonian $H_2 \sim L_E \mathcal{H}_U(0)$.  Upon reduction of the energy band width, $\Lambda_0\sim v_F/\xi \to \Lambda=\Lambda_0 e^{-l}$, new operators are generated in the effective Hamiltonian of the barrier. They have the form of higher powers of $H_2$. The most relevant of those is $H_4 \sim \Lambda r_4 \cos{\left(4\Phi_{c+}(0)\right)}$, with the scaling dimension $4K_{c+}-1$.  The correction $\delta G^{(4)}$ can be obtained by the lowest  order perturbation theory in $H_4$, $\delta G^{(4)} \sim (e^2/h)\int t dt \langle [H_4(t), H_4(0)]\rangle \propto T^{8K_{c+}-2}$.

For $K_{c+} < 1/4$ the $H_4$ is relevant and $\delta G^{(4)}$ diverges at zero temperature. The RG fixed point (FP) at perfect transmission is unstable, and the system flows to strong backscattering even at $L_E\ll \zeta$. We show below that the strong backscattering FP is stable. The expected RG flow is as shown in Fig.~\ref{fig:pnjunction}(d). It corresponds to vanishing zero temperature conductance. We note that the temperature dependence of conductance  is nonmonotonic. In the regime of applicability of  perturbation theory $\delta G= - a \left( L_E/\zeta\right)^2 \left(T/\Delta \right)^{2K_{c+}} - b \left( L_E/\zeta\right)^4 \left(T/\Delta \right)^{8K_{c+}-2}$, where $a$ and $b$ are constants on the order of $e^2/h$. The maximum  conductance is reached at $T\sim \Delta \left( L_E/\zeta\right)^{1/(1-3K_{c+})} $.

Because of the large value of the interaction constant, $e^2/ v_F \approx 2.7$ and the sensitivity of the forward scattering matrix element $V(0)$ to screening by the gates both cases  $K_{c+}<1/4$ and $K_{c+}>1/4$ may  be realized.

For $K_{c+}>1/4$ all backscattering operators generated in the process of renormalization are irrelevant. Therefore the FP  at perfect transmission is stable for weak backscattering, $L_E\ll \zeta$. At the same time, the analysis below shows that a perfect reflection fixed point is stable at $K_{c+}<1$. This indicates the existence at $1/4<K_{c+}<1$ of a quantum phase transition in the conductance of pn-junction, controlled by the ratio $L_E/\zeta$.

At strong backscattering, $L_E\gg \zeta$, the pn-junction may be viewed as two semi-infinite Luttinger liquids separated by a strong barrier where the bosonic fields are pinned to one of the classical minima of the Umklapp Hamiltonian. The minima form a periodic lattice in the four-dimensional space
$\Phi_j=\pi n_j/2$, where $n_j$ are integers, which are either all even or all odd. Charge transport between the p- and n- regions proceeds via tunneling between different points in the lattice~\cite{Yi1998}. The tunneling operator corresponding to the shift $\{ n_j \} \to \{ n_i+ \delta n_j\}$ is $ \exp\{i\sum_j [\theta_j(x_1)-\theta_j(x_2)]\delta n_j/2 ]\}$, where $x_1$ and $x_2$ label the points just to the left and to the right of the barrier. Its scaling dimension is $1 -\sum_j (\delta n_j)^2/(4 K_j)$. The operators that transfer charge across the junction must have nonzero $\delta n_{c+}$. For $K_{c+}<1$ all of them are irrelevant, and thus the zero temperature conductance vanishes.

The stable FPs at perfect transmission and  reflection for $1/4<K_{c+}<1$ must be separated by an unstable FP with an intermediate value  of zero temperature conductance. The pn-junction can be tuned to this FP by adjusting the parameter $L_E$. The dependence of conductance on $L_E$ at low temperature is schematically presented in Fig.~\ref{fig:pnjunction}(c). The critical conductance and critical exponents at the intermediate fixed point depend on the values of the Luttinger parameters $K_j$.

To analyze the intermediate FP near $K_{c+}=1/4$ we note that only two of the operators generated in the course of renormalization have scaling dimensions that are reasonably small: $H_4$ and $H_U$ (respective scaling dimensions $4K_{c+}-1$ and $K_{c+}$). All other operators are strongly irrelevant. The proximity of the values of $K_{c+}$ corresponding to marginality of $H_U$ and $H_4$ ($K_{c+}=0$, and $K_{c+}=1/4$) enables one to use a kind of $\epsilon$-expansion in which both $\epsilon =K_{c+}$ and $\epsilon_4=K_{c+}-1/4$ are assumed  small. In this case the intermediate FP is perturbatively accessible.
Writing the effective pinning Hamiltonian as $H_{eff}=\Lambda [r_4 \cos(4\Phi_{c+}) + \cos(2\Phi_{c+})\sum_a r_a\cos(2\Phi_a)]$, where $ r_4 , \, r_{a}$
are dimensionless amplitudes and $a$ labels the neutral modes, we obtain the RG equations to leading order in $\epsilon, \epsilon_4$,
\begin{subequations}
\label{eq:RG_epsilon_4}
\begin{eqnarray}
  d_l r_a &=& -\epsilon \, r_a+ r_a r_4 ,\\
  d_l r_4&=&-4\epsilon_4 r_4+\sum_a  r^2_a/4,
\end{eqnarray}
\end{subequations}
where $d_l= d/dl$. The intermediate FP exists for $\epsilon_4>0$ and is located at $r^*_4=\epsilon$ and $\sum_a r^{*2}_a=16\epsilon_4 \epsilon $. The FP conductance can be found by using lowest order perturbation theory with FP values of the reflection amplitudes~\cite{Kane1992}, $G^*=4 (e^2/h)(1- \epsilon_4 -\epsilon^2/32\epsilon_4)$.  The presence of the $\epsilon_4$ in the denominator in the last term signals that Eq.~(\ref{eq:RG_epsilon_4}) is valid provided $\epsilon_4 \gg \epsilon^2$. Near the FP the conductance  behaves as $G(T)-G^*\sim (e^2/h)(1-L_E/L_E^*)(\hbar v_F/T\xi)^{\lambda_+}$ where $\lambda_+=-2\epsilon_4+2\sqrt{\epsilon^2_4+2\epsilon\epsilon_4}$.
A similar analysis in the limiting case of $1-K_j =\eta \to 0_+$ gives the intermediate FP conductance $G^* \propto \eta^2$. Interpolation between these two limiting cases gives
the RG flow shown in Fig.~\ref{fig:pnjunction}(d).

The above picture is modified if single-electron backscattering is present. In a symmetric armchair CNT it can be caused by a magnetic field $B$ applied along the CNT axis or by the electron-phonon (ep) interactions. The corresponding backscattering Hamiltonian may be written as~\cite{Ajiki, Kane1998, Chen2008}
\begin{equation}\label{eq:single_electron_backscattering}
\delta H= -i \int dx \sum_{\alpha r \sigma}\psi^{\dagger}_{\alpha r \sigma}(x)
\psi_{\alpha -r \sigma} r [\alpha\Delta_u(x) +\Delta_B],
\end{equation}
where $\Delta_B$ and $\Delta_u$ are respectively the gaps induced in the single particle spectrum  by the magnetic field, $\Delta_B=(\pi/2)e B v_F R $, and by the lattice deformation. At low temperatures only the twist acoustic (TA) phonons are important and $\Delta_u(x)=g_T \partial_x u(x)$, where $u(x)$ is the TA mode  displacement and $g_T\sim 1/\sqrt{N} \ll 1$ is the corresponding coupling constant. As in the case of two-particle processes, single electron backscattering is effective only in the vicinity of the pn-junction, $|x|\lesssim L_E$. The  zero-transmission FP is stable with respect to single electron backscattering.

For $L_E\ll \zeta$ and at small $B$ and $g_T$ the stability of the FP at perfect transmission with respect to single particle backscattering may be examined by perturbation theory. Bosonizing the fermion operators in Eq.~(\ref{eq:single_electron_backscattering}) as in Eq.~(\ref{eq:bosonization_of fermions}) we find the lowest order correction $\delta G \sim -(e^2/h)[(L_E/\xi)^2 (g_T^2/\rho s_T^3) (\xi T/v_F)^{(3+K_{c+})/2} +(L_E \Delta_B/v_F)^2 (v_F T/\xi)^{(K_{c+}-1)/2}]$. Here $\rho$ is the mass per unit length of the CNT and $s_T$ is the speed of sound for the TA mode~\cite{Chen2008}.
The temperature dependence of the two terms here shows the stability of the perfect transmission FP with respect to the ep backscattering and its instability with respect to $B\neq 0$.

Combining the stability analysis of the perfect transmission and perfect reflection FPs we conclude that the intermediate FP is not destroyed by the ep interactions. However application of a magnetic field drives the zero temperature conductance of the junction to zero.
This reveals that magnetoresistance of a CNT pn-junction~\cite{Andreev2007} is strongly enhanced by the Luttinger liquid effects.

In conclusion, the Mott-insulating state of an armchair CNT should manifest itself in the dependence of the pn junction conductance $G$ on the temperature $T$ and built-in electric field $E$. In a broad range of parameters, the $G(T,E)$ dependence is controlled by a $T=0$ quantum phase transition occurring at some value of $E=E^*$. The critical behavior of $G(T,E)$ is controlled by an unstable fixed point with  conductance $G^*<4e^2/h$. That suggests the possibility of the scaling analysis of experimental data used in investigations of quantum phase transitions~\cite{QPTReview}.

This work was supported by the DOE grant DE-FG02-07ER46452 (W.C. and A.V.A) and the NSF DMR Grant No. 0906498 at Yale University (L.I.G.), and the Nanosciences Foundation at Grenoble, France. The hospitality of CEA Grenoble (France) is greatly appreciated.

\end{document}